\documentclass[aps,pra,twocolumn,amssymb,superscriptaddress,nofootinbib,longbibliography]{revtex4-2}
\pdfoutput=1
\usepackage[utf8]{inputenc}
\usepackage[english]{babel}
\usepackage[T1]{fontenc}
\usepackage{tikz}

\usepackage{graphicx}
\usepackage{color}
\usepackage{braket}
\usepackage{verbatim}
\usepackage{amsmath}
\usepackage{amsthm}
\usepackage{amsfonts}
\usepackage{mathtools}
\usepackage{braket}
\usepackage{float}
\usepackage{hyperref}
\usepackage{lipsum}
\usepackage{bbold}
\usepackage{dsfont}
\newtheorem{theorem}{Theorem}
\newtheorem{lemma}[theorem]{Lemma}

\usepackage[toc,page,header]{appendix}
\usepackage{makecell}

\begin{document}

\title{Generating graph states with a single quantum emitter and the minimum number of fusions}
\author{Matthias C. L\"{o}bl}
\affiliation{Center for Hybrid Quantum Networks (Hy-Q), The Niels Bohr Institute, University of Copenhagen, Blegdamsvej 17, DK-2100 Copenhagen {\O}, Denmark}
\author{Love A. Pettersson}
\affiliation{Center for Hybrid Quantum Networks (Hy-Q), The Niels Bohr Institute, University of Copenhagen, Blegdamsvej 17, DK-2100 Copenhagen {\O}, Denmark}
\author{Andrew Jena}
\affiliation{Department of Combinatorics and Optimization, University of Waterloo and Institute for Quantum Computing, University of Waterloo}
\author{Luca Dellantonio}
\affiliation{Department of Physics and Astronomy, University of Exeter, Stocker Road, Exeter EX4 4QL, United Kingdom}
\author{Stefano Paesani}
\affiliation{Center for Hybrid Quantum Networks (Hy-Q), The Niels Bohr Institute, University of Copenhagen, Blegdamsvej 17, DK-2100 Copenhagen {\O}, Denmark}
\affiliation{NNF Quantum Computing Programme, Niels Bohr Institute, University of Copenhagen, Blegdamsvej 17, DK-2100 Copenhagen {\O}, Denmark.}
\author{Anders S. S\o{}rensen}
\affiliation{Center for Hybrid Quantum Networks (Hy-Q), The Niels Bohr Institute, University of Copenhagen, Blegdamsvej 17, DK-2100 Copenhagen {\O}, Denmark}

\begin{abstract}
Graph states are the key resources for measurement- and fusion-based quantum computing with photons, yet their creation is experimentally challenging. We optimize a hybrid graph-state generation scheme using a single quantum emitter and linear optics Bell-state measurements called fusions. We first generate a restricted class of states from a single quantum emitter and then apply fusions to create a target graph state, where we use a dynamic programming approach to find the construction that requires the lowest possible number of fusions. Our analysis yields a lookup table for constructing $\sim 2.8\times 10^7$ non-isomorphic graph states with the minimum number of fusions. The lookup table covers all graph states with up to eight qubits and several other ones with up to 14 qubits. We present construction protocols of selected graph states and provide the lookup table. For large graph states that are not in the lookup table, we derive bounds for the required number of fusions using graph-theoretic properties. Finally, we use the lookup table to search for the best graph codes for loss-tolerant encodings, given a fixed number of fusions for their construction.
\end{abstract}

\maketitle

\section{Introduction}

Measurement- and fusion-based quantum computing paradigms are particularly promising for photonic qubit platforms~\cite{Raussendorf2001, Bartolucci2021}. In both approaches, quantum computing is performed via measurements consuming qubits from a set of entangled resource states, so-called graph states~\cite{Hein2004, Hein2006}. What so far has hindered quantum computing with photonic graph states is photon loss and the difficulty of realizing entanglement between photons. Small graph states can be created in a heralded way using single photons, linear optics, and measurements~\cite{Bartolucci2021b, Maring2024}. However, heralded graph-state generation is a probabilistic process and the probability of success becomes extremely small for an increasing number of qubits.

To mitigate this scalability issue, schemes using a single quantum emitter with a spin degree of freedom have been suggested for deterministically generating graph states~\cite{Lindner2009, Gheri1998, Tiurev2021} and implemented~\cite{Thomas2022, Coste2022, Cogan2023, Meng2023, Huet2024}. A limitation is that the class of graph states that can be generated with this method is restricted to states with a one-dimensional entanglement structure. This class of states does not include many important graph states with applications in quantum communication and fault-tolerant quantum computing~\cite{Raussendorf2006, Azuma2015, Bell2022}. To enlarge the family of graph states that can be generated, interacting quantum emitters, quantum emitters coupled to static ancilla qubits~\cite{Economou2010, Li2022, Pettersson2024}, or delayed feedback~\cite{Pichler2017, Zhan2020, Wan2021, Ferreira2024} can be employed.

However, corresponding experimental realizations are challenging. Certain quantum dots are, for instance, excellent quantum emitters~\cite{Warburton2013, Ding2023}, and coupling between different quantum dots has been demonstrated~\cite{Tiranov2023} as well as coupling to a static ancilla~\cite{Appel2024}. The issue is that high-fidelity realizations of these operations remain elusive, making it difficult to exploit them for resource state generation. Alternatively, color centers provide static ancilla qubits in the form of nuclear spins~\cite{Parker2024, Bradley2019, Stas2022}. However, the coupling time to the ancilla qubits is slow, their number is limited, and the photonic properties of color centers are typically worse than those of quantum dots~\cite{Ding2023}. These limitations have motivated minimizing the required number of quantum emitters, static ancilla qubits, and emitter-emitter gates participating in the graph-state generation~\cite{Li2022, Ghanbari2024, Kaur2024}. An alternative approach based on delayed feedback~\cite{Pichler2017} has been realized in superconducting circuits~\cite{Ferreira2024}. However, this approach is very demanding for optical photons since photons need to be coupled into the photon source hosting the quantum emitter at a later point in time. Given that the best in-/out-coupling efficiencies of photon sources in the optical regime are currently at about $0.7$~\cite{Ding2023}, delayed feedback techniques may remain inefficient in the near future.

To avoid these issues, we consider a hybrid strategy using only a single quantum emitter containing a single internal spin qubit and linear optics type-II fusions~\cite{Browne2005, Lee2023, Lobl2023}. We first generate an initial graph state using the quantum emitter~\cite{Lindner2009} and then apply local gates and fusions to construct a target graph state that a single quantum emitter cannot generate~\cite{Hilaire2023}. Since fusions only succeed with a finite probability and destructively measure two photonic qubits~\cite{Browne2005, Gimeno2016}, minimizing the number of fusion operations is important for such an approach. The fewer fusions, the higher the probability that the targeted graph states are successfully generated and the fewer photonic qubits need to be generated.

Our main contribution in this work is to find optimal graph state constructions, where optimal means that only the minimum possible number of successful fusions is required. We provide a lookup table for many such graph state constructions where fusions are applied to an initial state that a single quantum emitter can generate\footnote{A similar optimization is minimizing the number of CZ-gates in a circuit-based construction of graph states~\cite{Kumabe2024, Jena2024}. The main difference is that our graph state constructions do not start from an empty graph but from an arbitrary \textit{caterpillar tree}~\cite{Harary1973} graph state that already contains entanglement.}. The key to computing such a lookup table is a dynamic programming approach. For larger graph states that are not part of the lookup table, we give bounds for the required number of fusions in graph-theoretical terms. Furthermore, we use the obtained lookup table to find optimal constructions of several interesting graph states such as repeater graph states~\cite{Azuma2015}. Finally, we use the search table to identify graph codes with high photon loss thresholds~\cite{Bell2022} that require only a few fusions to generate them. 

\section{Graph states, optical quantum emitters with a spin, and fusions}
\begin{figure*}[!t]
\includegraphics[width=1.0\textwidth]{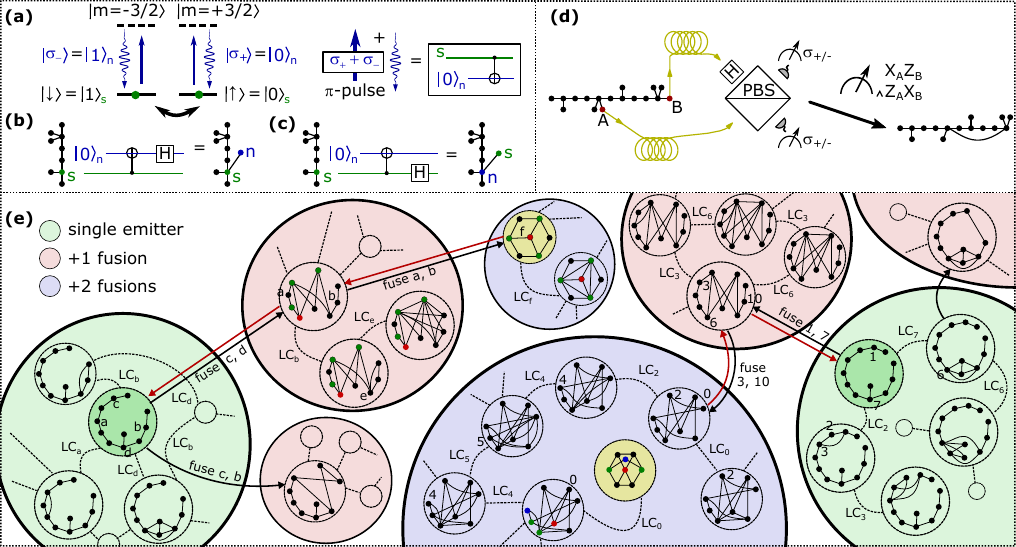}
\caption{\label{fig:lookup} Generating graph states with a single quantum emitter and a minimum number of fusions. \textbf{(a)} Exemplary level structure of a single quantum emitter with two spin ground states ($\ket{\uparrow}, \ket{\downarrow}$). When a linearly polarized $\pi$-pulse (blue) simultaneously excites the two optical transitions, the subsequent decay via photon emission realizes a spin-photon CNOT gate between the new emitted photon (n) and the spin (s)~\cite{Lindner2009}. \textbf{(b)} A $\pi$-pulse followed by an $H$-gate (Hadamard) on the new photon attaches the new photon as a \textit{leaf} to the graph state. \textbf{(c)} Instead, applying the $H$-gate on the spin makes the spin the leaf qubit. The combination of (b, c) enables generating graph states with a caterpillar~\cite{Harary1973} structure. \textbf{(d)} Two photons $A, B$ are sent to a fusion setup made of a polarizing beam splitter (PBS), two circular-polarization-resolving detectors, and an $H$-gate. When the two photons are detected by two different detectors, the setup measures the parities $X_AZ_B\land Z_AX_B$ (fusion success), generating a new graph edge. \textbf{(e)} Illustration of the lookup table. Within graph orbits (big circles), the dashed edges between the graphs represent single-qubit gates corresponding to \textit{local complementations}~\cite{Hein2004, Hein2006} (LC). The colors of the graph orbits represent the required number of successful fusions. The solid black arrows represent fusions that transform graph states into new states belonging to different graph orbits. Therefore, fusions represent links connecting graph orbits into an \textit{orbit graph}. A protocol for constructing target graph states (examples in yellow) is found by traversing the orbit graph backward (red arrows) and applying LCs to navigate within the graph orbits. Some graph-state nodes are colored to indicate which node is which upon fusion or LC.}
\end{figure*}
In this section, we introduce the basic concepts of graph states and fusions and define the graph-state generation scheme. For the latter, we employ a single quantum emitter with a spin.

\textit{Graph states $-$} In a graph state, qubits are represented by graph vertices $V$ and a controlled-$Z$ gate $CZ_{ij}$ between the qubits $i, j$ is represented by an edge $(i,j)\in E$~\cite{Hein2006}. Given the graph $G=(V,E)$, the corresponding graph state is $\ket{G}=\prod_{(i,j)\in E} CZ_{ij}\ket{+}^{\otimes{V}}$ with $\ket{+}=\frac{1}{\sqrt{2}}(\ket{0}+\ket{1})$\footnote{The graph state is well-defined since the gates $CZ_{ij}=\left(\ket{00}\bra{00}+\ket{01}\bra{01}+\ket{10}\bra{10}-\ket{11}\bra{11}\right)_{ij}$ commute.}. Graph states are stabilizer states meaning that a graph state $\ket{G}$ is the unique eigenstate with $+1$ eigenvalues of the stabilizer generators $\{S_i=X_i\prod_{j\in N(i)} Z_j\}$, where $i\in V$, $N(i)$ are all neighbors of graph node $i$, and $X=\ket{0}\bra{1}+\ket{1}\bra{0}$ and $Z=\ket{0}\bra{0}-\ket{1}\bra{1}$ are Pauli matrices~\cite{Hein2004, Hein2006}. Different graph states can be \textit{local Clifford equivalent}, meaning that one state can be converted into the other by local Clifford gates. Two graph states are Clifford equivalent if and only if one graph can be converted into the other by \textit{local graph complementations}~\cite{Nest2004, Hein2006}. We call the set of all graph states that are local Clifford equivalent a \textit{graph orbit}~\cite{Adcock2020, Cabello2011}.

\textit{Quantum emitters $-$ } A single optical quantum emitter with a spin can be used to generate a photonic graph state with a graph structure of a caterpillar tree~\cite{Lindner2009, Pettersson2024}. \textit{Caterpillar trees} are tree graphs with a central path to which all other vertices are connected~\cite{Harary1973} (we will often use the terms \textit{caterpillar tree} or \textit{caterpillar} interchangeably with \textit{graph state with a graph structure of a caterpillar tree}).

A suitable level scheme for the quantum emitter is shown in Fig.~\ref{fig:lookup}(a). The level structure has two optically excited states with spin quantum numbers $m=\pm\frac{3}{2}$ where the degeneracy with the $m=\pm\frac{1}{2}$ excited states is lifted. First, the static spin qubit, $s$, is initialized in the superposition state $\ket{+}_s=\frac{1}{\sqrt{2}}(\ket{\uparrow}+\ket{\downarrow})$ corresponding to a graph state with a single node. Applying a linearly polarized ($\sigma_++\sigma_-$) optical $\pi$-pulse simultaneously excites the two optical transitions to two excited levels with angular momentum quantum numbers $m=\pm  3/2$. For a spin initially in the state $\ket{\uparrow}=\ket{0}_s$, the photon that is emitted upon decay from the corresponding exited state is in the state $\ket{\sigma_+}=\ket{0}_n$ (circularly $\sigma_+$-polarized) and, for a spin in the state $\ket{\downarrow}=\ket{1}_s$ the photon is in the state $\ket{\sigma_-}=\ket{1}_n$ (circularly $\sigma_-$-polarized). This leads to the state $\frac{1}{\sqrt{2}}(\ket{0}_s\ket{0}_n+\ket{1}_s\ket{1}_n)$. Therefore, the optical pulse with subsequent photon emission implements a CNOT$_{s,n}$ gate between the spin and the new photonic qubit (see Fig.~\ref{fig:lookup}(a), right), providing the key source for the entanglement generation. Applying a single-qubit Hadamard gate $H=\frac{1}{\sqrt{2}}(X+Z)$ to the new photonic qubit turns the state into a graph state where the new photon is attached as a leaf connected to the spin. The same applies if the spin was part of a larger graph state before, as shown in Fig.~\ref{fig:lookup}(b). Alternatively, applying the $H$-gate to the spin makes the spin a leaf qubit of the caterpillar graph, with the new photon taking the place of the spin, as shown in Fig.~\ref{fig:lookup}(c). Repeating this process allows the deterministic growth of a photonic graph state with any caterpillar structure~\cite{Lindner2009} (see Lemma~\ref{lemma_caterpillar} in Appendix~\ref{sec_one_emitter}).

\textit{Fusions $-$ } Fusions are probabilistic Bell state measurements consuming the measured photonic qubits. Fusion success generates additional entanglement and fusions can thus be used to create graph states that a single quantum emitter cannot generate. Without the use of ancillary photons, a fusion succeeds with a probability of $0.5$~\cite{Browne2005}\footnote{The success probability can be boosted to $p_s>0.5$ with ancilla photons~\cite{Grice2011} which, however, increases the chance of photon loss.}, resulting in the simultaneous measurement of two two-qubit operators from the Pauli group~\cite{Lobl2024}. In turn, fusion failure corresponds to two single-qubit Pauli measurements. We consider type-II fusions~\cite{Browne2005, Lobl2024} since, for this type of fusion, fusion success/failure and photon loss are heralded~\cite{Gimeno2016} by the detection pattern of the fusion setup. In the absence of errors, it will therefore always be known whether the desired graph state has been constructed successfully. The graph transformation corresponding to a successful fusion depends on the exact linear optics setup that determines which two-qubit operators are measured upon fusion success~\cite{Lobl2024}. Fig.~\ref{fig:lookup}(d) shows a setup for a fusion that measures the two parities $X_AZ_B\land Z_AX_B$ upon fusion success, where $A, B$ are the fusion photons that are destructively measured. This fusion success case adds an additional edge to the graph state as illustrated on the right of Fig.~\ref{fig:lookup}(d)\footnote{The graph transformation may involve additional local Pauli gates. We neglect these Pauli gates as they depend on the probabilistic measurement result and correspond to stabilizer signs, which can be tracked independently~\cite{Gottesman1998, Aaronson2004}.}. A quantum optical explanation of similar setups can be found in Ref.~\cite{Gimeno2016}.

\section{Computing the lookup table}
Now, we describe how we compute a lookup table for constructing a target graph state from caterpillar tree graph states and a minimum number of fusions. To build a desired graph state, a possible approach (method 0) would be looking for all parent graph states from where the target state can be reached by a single fusion. For all these parent states, one could proceed in the same way until a graph is reached that a single quantum emitter can generate. The main issue with method 0 is that the number of possible inverse fusions and so the number of parent graph states grows more than exponentially with the graph-state size. Furthermore, local gates can be applied before fusions, so it would be necessary to apply inverse fusions to all graph states within the same local equivalence class (graph orbit).

Instead, we use a dynamic programming technique~\cite{Bellman1965} that approaches the problem from the other end. First, we generate all graph states up to a certain number of qubits that are caterpillar trees or local Clifford equivalent to them (same graph orbit). Second, we take these graph states as parent graph states and apply fusions to all of them, generating new graph states and new graph orbits. The second step is repeatedly applied until no further graph states can be found. This is a dynamic programming approach as it avoids computing identical graph states several times when two different graph states share part of their construction (overlapping sub-problems). Therefore, this approach strongly reduces the computational cost compared to applying method 0 to every graph state. Furthermore, the obtained constructions are optimal in the number of fusions. In fact, the minimum number of fusions required to generate a graph $G$ is given by the minimum number of fusions over all parent graph states of $G$ plus one fusion more. This optimal sub-structure is expressed by the following functional equation: $f(G)=\text{min}_{G', G' > G}\,\,f(G')+1$, where $f(G)$ is the function that computes the minimum number of fusions to build the graph state $G$ and $G' > G$ indicates that $G$ can be obtained by applying a fusion to $G'$. A very similar method has originally been suggested for finding optimal moves in endgames of two-party games such as chess~\cite{Bellman1965}, where the resulting lookup table is today known as a tablebase.

\textit{Implementation $-$ }We start by generating all graph states with a caterpillar structure up to 14 qubits, including sets of detached caterpillars (these graph states can be obtained from a caterpillar tree by applying $Z$-basis measurements). For every caterpillar, we also add all its local Clifford equivalent states by applying local graph complementations~\cite{Hein2004, Hein2006} (see Appendix~\ref{sec:LC}). In this way, we obtain the initial set of graph states, which is illustrated by the green-shaded graph orbits in Fig.~\ref{fig:lookup}(e). Among all graph states in one of these graph orbits, there is always one caterpillar graph that is highlighted in dark green. All these initial graph states can be generated by a single quantum emitter plus local gates~\cite{Lindner2009, Paesani2023}. Furthermore, we show that no additional graph states can be generated with these resources (see Theorem~\ref{theorem_only_caterpillar} in Appendix~\ref{sec_one_emitter}), justifying the choice of the initial graph states. We consider isomorphic\footnote{Two graphs are isomorphic when one can be converted into the other by relabelling the qubits correspondingly.} graphs the same, storing just one out of several isomorphic graphs in the lookup table.

In the next step, more graph states are obtained by applying fusions to existing graph states. We show in Appendix~\ref{sec_justify} that it is sufficient to consider a single type of fusion (see Theorem~\ref{theorem_one_fusion}). The reason is as follows: assume that applying a successful fusion of a certain type to a graph state $\ket{G_1}$ yields a state $\ket{G_2}$. When applying a different type of successful fusion, there always exists a graph state $\ket{G_1'}$ in the orbit of $\ket{G_1}$ that becomes a state that is local Clifford equivalent to $\ket{G_2}$. Therefore, we only apply fusions that, upon success, measure the parity pair $X_AZ_B\land Z_AX_B$ between qubits $A, B$\footnote{To obtain the post-fusion states, we apply the graph transformation rules from Ref.~\cite{Lobl2024} but alternative methods such as ZX-calculus could be used instead~\cite{Felice2024}. Choosing the parity measurement $X_AZ_B\land Z_AX_B$ is convenient since it transforms a graph state to a graph state without additional Clifford gates (if $A, B$ are not connected in the initial graph state)~\cite{Lobl2024}. In our final lookup table, there are only $34$ cases where two connected qubits are fused. The corresponding local Clifford gates that transform the post-fusion state into a graph state are found upon lookup using the graph transformation rules from~\cite{Lobl2024}.}.

For all graph states in the initial graph orbits (green-shaded in Fig.~\ref{fig:lookup}(e)), we apply fusions between all pairs of graph nodes. If a fusion applied to $G_1$ generates a graph state $G_2$ that does not yet exist in the lookup table, we add a new graph orbit with $G_2$ and its local Clifford equivalent states. The applied fusion is stored as a link from the graph orbit of $G_1$ to the graph orbit of $G_2$ (black arrows in Fig.~\ref{fig:lookup}(e)). Graph orbits that exist already are not generated again, and also, no additional link to them is stored. In this way, we obtain all graph orbits that can be generated by a single fusion (shaded red in Fig.~\ref{fig:lookup}(e)).

In this procedure, we need to test for every graph state generated by a fusion whether it is already present in the lookup table. To this end, we first compute its Weisfeiler-Lehman graph hash, which is fast~\cite{Shervashidze2011}, and check if there already exists a graph with the same hash in a corresponding dictionary. If the answer is no, the graph is new\footnote{The Weisfeiler-Lehman graph hash is identical for isomorphic graphs, yet there can be hash-collisions where two non-isomorphic graphs yield the same hash~\cite{Shervashidze2011}. Therefore, we store a list of graphs per hash, which still gives an efficient lookup table for graphs since the hash collisions are very rare and the length of the lists are therefore small.}. Only if the answer is yes, we perform slower true isomorphism tests~\cite{Cordella2001} with the existing graphs that have the same hash.

Once all graph orbits that can be reached by a single fusion are generated, we proceed by applying another fusion to all graph states in these new graph orbits. In this way, we obtain the graph orbits that require two successful fusions (blue-shaded in Fig.~\ref{fig:lookup}(e)). This procedure is repeated until no new graph orbits are generated. This way, we obtain an \textit{orbit graph $\mathcal{G}$} where nodes represent graph orbits and fusions represent edges connecting them, respectively. Since we only store the fusion from the orbit of $G_1$ to the orbit of $G_2$ if the latter orbit did not exist before, there is at most one incoming edge (fusion) per orbit and $\mathcal{G}$ has a tree structure. Therefore, the lookup table always provides one optimal graph-state construction (alternative constructions via other graph orbits and the same number of fusions can exist). An illustration of the orbit graph $\mathcal{G}$ is shown in Fig.~\ref{fig:lookup}(e).

Once $\mathcal{G}$ is computed, it can be used as a lookup table for optimal constructions of desired target graph states\footnote{The number of required successful fusions is minimum (i.e. the construction is optimal) if only local Clifford gates can be applied. Although local unitary (LU) equivalence equals local Clifford (LC) equivalence for small graph states~\cite{Hein2006, Cabello2009, Burchardt2024}, it is generally not identical~\cite{Ji2007, Tsimakuridze2017}. Whether using non-Clifford gates is an additional resource is an open question. Given that the smallest known counterexamples of LU-LC equivalence are 27-qubit graph states~\cite{Ji2007, Tsimakuridze2017}, the gain of using non-Clifford gates is presumably modest or absent for the considered graph state sizes~\cite{Burchardt2024}. Non-Clifford gates could be integrated by using generalized local complementations~\cite{Claudet2024}.}. First, we find the orbit containing the graph state (highlighted in yellow in Fig.~\ref{fig:lookup}(e)). This is done by using a dictionary embedded in the lookup table that gives the graph orbit (value) for a given index labeling the target graph. The unique index that labels a graph can be efficiently found by using its Weisfeiler-Lehman graph hash~\cite{Shervashidze2011} to navigate to the position in the data structure where the graph is stored. Second, we look at the incoming link (fusion) that connects the current orbit to its parent, i.e, the orbit from where the current one can be reached by a single fusion. This is illustrated by the solid red arrows in Fig.~\ref{fig:lookup}(e). The procedure is repeated until an orbit is reached that can be generated by a single quantum emitter. The necessary fusions to reach the desired graph states are then given by listing the traversed links (solid red arrows in Fig.~\ref{fig:lookup}(e)) between the graph orbits in inverse order.

Every fusion link connects two different states in two different graph orbits by a fusion, but a sequence of LCs (dashed lines in Fig.~\ref{fig:lookup}(e)) might be required to navigate to the graph $G$ that is connected to the parent orbit by an incoming link. We find these Clifford gates by repeatedly applying LCs to the current graph and traversing the graph orbit breadth-first until we arrive at $G$ (done for every lookup). This is an inefficient method and currently the main contribution to the overall time required to look up a graph state, but it can be significantly improved using, e.g., the algorithm from Refs.~\cite{Nest2004b, Bouchet1991}. However, for any node number between $1-14$, the overall lookup time of a graph state construction is on average below $3\ $s and is much less for below 10 nodes\footnote{For arbitrary 14-qubit graph states, this would likely be much slower. However, the considered 14-qubit states are local Clifford equivalent to caterpillar trees, and therefore, the corresponding graph obits are particularly small.}. This is sufficiently fast for a relatively smooth user experience; therefore, we have not improved the algorithm in the current implementation. We provide a program that uses the pre-computed lookup table and automatically finds the exact sequence of LCs and fusions to construct an arbitrary target graph state~\cite{git2024}.

\section{Results}
\begin{figure*}[!t]
\includegraphics[width=1.0\textwidth]{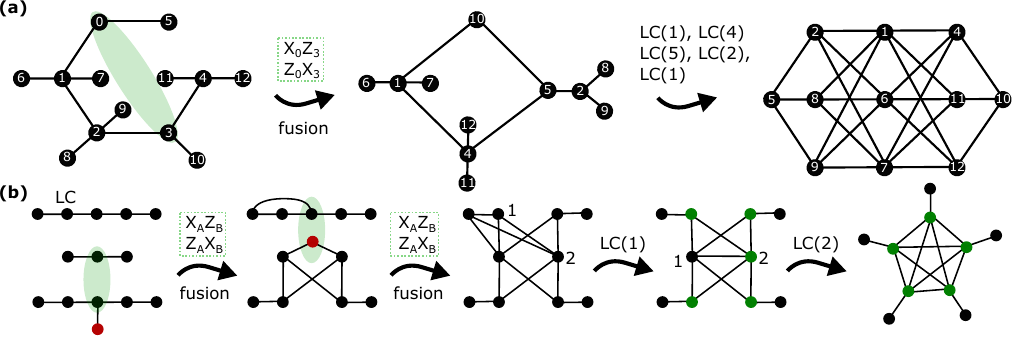}
\caption{\label{fig:examples} Two exemplary graph state constructions. Every construction starts from a state that a single quantum emitter can create deterministically and uses the minimum number of fusions. For both states, two interacting static qubits would be required to generate them deterministically without any fusion~\cite{Li2022}. \textbf{(a)} Construction of an $11$-qubit crazy graph~\cite{Morley2019} with one fusion. \textbf{(b)} Construction of a $10$-qubit repeater graph state with two fusions. The colors indicate which node is which before and after a fusion or local complementation.}
\end{figure*}
\subsection{Graph state constructions}
Having computed the lookup table, we can use it to find optimum graph-state constructions with a minimum number of successful fusions. In Fig.~\ref{fig:examples}, we show two examples of constructions of graph states that we can generate from a single quantum emitter and a minimum number of fusions. The presented graph states have applications, e.g., in quantum communication. In Fig.~\ref{fig:examples}(a), we show the construction of a graph state for loss tolerant quantum teleportantion~\cite{Morley2019} and, in Fig.~\ref{fig:examples}(b), we show the construction of a repeater graph state~\cite{Azuma2015}.

Since a standard linear-optics fusion succeeds with a probability of $p_s=0.5$~\cite{Browne2005, Lobl2024}, an $n$-fusion construction only gives the desired state with a success probability of $1/2^n$. Therefore, a low number of fusions, e.g., one or two like in Fig.~\ref{fig:examples}, is highly desirable\footnote{The presented construction of the $N=5$ repeater graph state in Fig.~\ref{fig:examples}(b) requires one fusion less than previously proposed constructions using a single quantum emitter~\cite{Buterakos2017} (the same applies to the $N=4$ repeater graph state, see Appendix~\ref{sec_constr}).}. For the presented graph-state constructions that require only a few fusions, the corresponding average graph-state-generation rate could still be tens of MHz when using fast quantum emitters such as quantum dots~\cite{Cogan2023, Meng2023, Huet2024}. In comparison, a simple construction from star-shaped resource states~\cite{Lobl2023, Lobl2024b} with every graph edge being generated by a single fusion would require way more fusions ($24$ and $15$ for the cases in Fig.~\ref{fig:examples}(a,b), respectively). The probability of successfully generating the graph state, and consequentially the average graph-state generation rate, would thus be extremely low even if several graph-state sources are multiplexed.

The required number of fusions can be reduced by adaptive strategies that do not repeat the entire protocol until, by coincidence, all fusions succeed at the same time. In particular, divide-and-conquer strategies can avoid the average number of consumed fusions scaling exponentially with the graph-state size~\cite{Barrett2005, Duan2005, Lee2023}. However, adaptive strategies have the disadvantage that photon loss and errors are amplified by storing parts of a graph state in a fiber while another part is still being generated. Furthermore, adaptive strategies may still require enormously many fusions when using small resource states (see Table 1 in Ref.~\cite{Lee2023}). Our graph state constructions use larger caterpillar graphs as resource states and thus require only very few successful fusions. In this case, repeating a non-adaptive protocol until all fusions succeed simultaneously is a reasonable strategy.

More graph state constructions can be found in Appendix~\ref{sec_constr} and the openly available lookup table (see Sec.~\ref{sec_data}). Although we restrict the size of the initial caterpillar graph states, all the found constructions are optimal regarding the number of fusions (see Theorem~\ref{theorem_optimality} in Appendix~\ref{sec_optimal}). However, there can be alternative constructions of a graph state that require the same number of fusions and move through different graph orbits.

\begin{figure}[!t]
\includegraphics[width=1.0\linewidth]{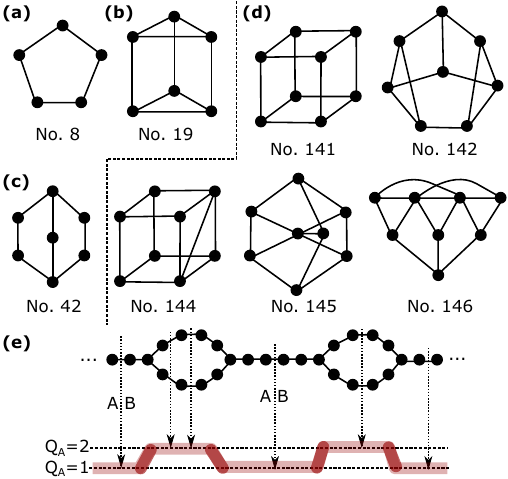}
\caption{\label{fig:many_fusion_graphs} Graph states that require particularly many fusions. Every graph represents a graph orbit, with the chosen graph being the one in the graph orbit that has the fewest edges. The numbers correspond to the graph orbit labels from Refs.~\cite{Hein2004, Cabello2009}. \textbf{(a)} Smallest graph state for which one fusion is required. \textbf{(b)} Smallest graph state for which two fusions are required. In the corresponding graph orbit, there is only one additional graph state that has a wheel structure~\cite{Looi2008}. \textbf{(c)} Smallest graph state for which $n_s^{min}<n_f^{min}+1$. The number of required static qubits to generate the graph state deterministically is $n_s^{min}=2$. Using one emitter only, one needs $n_f^{min}=2$ fusions. \textbf{(d)} The five eight-qubit graphs for which three fusions are required. The first graph is local Clifford equivalent to a wheel graph state~\cite{Looi2008}, and the second and fourth graphs correspond to graph codes from Ref.~\cite{Bell2022}. \textbf{(e)} Graph state for which $n_f^{min}\gg n_s^{min}=2$ when horizontally continuing the same pattern. On the bottom, we schematically illustrate the entanglement entropy $Q_A=Q_B$ (height function~\cite{Li2022}, red line) corresponding to a split into two subsets $A, B$ as indicated by the arrows. The required number of fusions $n_f^{min}$ equals the number of times $Q_A$ changes its value between one and two (dark red lines), saturating the bound from Eq.~\eqref{eq_climb}.}
\end{figure}

\subsection{Statistics of the lookup table}
The obtained lookup table contains all caterpillar graphs and detached caterpillars up to 14 qubits as well as all the states resulting after fusions (isomorphic graphs are stored as a single graph). The entire lookup table contains 27861431 graphs in 27034 graph orbits. We find 8843 graph orbits containing the initial caterpillars, and we find 16583, 1603, and 5 graph orbits that can be reached from the initial caterpillar graph states by a minimum number of one, two, and three fusions, respectively. When only counting connected graphs, the number of graph orbits is 10107, 1340, and 5 for one to three fusions, respectively. For zero fusions and connected graphs, we find $2144$ orbits. This number serves as a useful consistency check. For $n\geq 4$ nodes, there are $2^{n-4}+2^{\lfloor n/2-n \rfloor}$ non-isomorphic caterpillars~\cite{Harary1973}, and for one, two, and three nodes, there is only one caterpillar per node number. The overall number of caterpillars is therefore $\sum_{n=4}^{14}\left(2^{n-4}+2^{\lfloor n/2-n \rfloor}\right) + 3 = 2144$. At the same time, a caterpillar is a tree graph, and two locally equivalent tree graphs are isomorphic~\cite{Bouchet1988}. Therefore, every caterpillar makes its own graph orbit, and the expected number of initial (zero fusions) graph orbits is, again, $2144$.

The smallest graph state that requires at least one (two) fusion(s) is the five-qubit ring shown in Fig.~\ref{fig:many_fusion_graphs}(a) (the six-qubit graph state shown in Fig.~\ref{fig:many_fusion_graphs}(b)). The smallest graph states requiring three fusions have eight qubits. One graph state of every orbit is shown in Fig.~\ref{fig:many_fusion_graphs}(d). Interestingly, one of these states (the cube) has been found by a search for graph codes~\cite{Bell2022}. A possible correlation between the number of fusions required to build a graph state and the performance of this state as a graph code could be explored in the future. However, fewer fusions also enable the construction of very good graph codes (see subsection~\ref{sec_codes}).

To compare the number of obtained graph orbits to the number of all possible graph orbits (see Ref.~\cite{Cabello2011}), we filter the graph orbits corresponding to connected graph states. For up to eight qubits, we find that all possible graph orbits are contained in our lookup table (in agreement with Refs.~\cite{Hein2006, Cabello2011}, we find $1, 1, 2, 4, 11, 26, 101$ graph orbits for graphs with $2-8$ nodes, respectively). Since we do not find any construction with more than three fusions, this implies that all graph states up to eight qubits can be constructed by one quantum emitter and not more than three fusions. For nine (ten) qubits, we find that only $376$ ($1900$) of all possible $440$ ($3132$)~\cite{Cabello2011} graph orbits are contained in the lookup table, which means that about $14.5$ ($39.3$) percent of the nine(ten)-qubit graph orbits require more than two fusions\footnote{Note that nine or ten qubit states requiring more than two fusions do not appear in the lookup table since we restrict the initial states to 14 qubits.}.

\subsection{Bounds for the number of fusions}
\label{sec_bounds}
The provided lookup table gives optimum constructions of small graph states. For larger graph states, we give several bounds for the required number of fusions. First, we note that a lower bound for the minimum number of required fusions $n_f^{min}$ is given by the maximum of the corresponding number for all induced subgraphs (see Theorem~\ref{theorem_subgraph} in Appendix~\ref{sec_optimal}). This lower bound can be estimated by considering all subgraphs of the target graph that belong to the computed lookup table.

In the following, we give bounds for $n_f^{min}$ based on other properties of the target graph. We outline a connection between the required number of fusions and a different approach where photonic graph states are deterministically created using a quantum emitter with not only a single but multiple coupled static qubits~\cite{Li2022}. The minimum number of static qubits, $n_s^{min}$, for such a graph-state generation scheme can be determined with an argument based on bipartite entanglement entropy~\cite{Li2022}\footnote{In this approach, one assumes that arbitrary entangling gates can be applied between any pair of the $n_s$ static qubits. At least one of the static qubits must be a quantum emitter that can emit photons entangled with it.}.

Assume that all photons of the graph state are emitted in a certain order $p=p_1p_2...p_n$, where $p_i$ represents the index of the graph node/photon that is emitted in the $i$th time step and $p$ thus represents a permutation of graph nodes. Let $V_s$ be the set of static qubits and, at a certain point in time, $A=p_1p_2..p_k$ be all photons emitted in the past and $B=p_kp_{k+1}...p_n$ are all photons that will be emitted in the future. The entanglement entropy between $A$ and $B$ at a later point in time cannot exceed the entanglement entropy between $A$ and $V_s$ at the current point. Since the entanglement entropy cannot exceed the number of static qubits $|V_s|=n_s$, the entanglement entropy of $Q_A$ and $Q_B$ is bounded by $Q_A=Q_B\leq n_s$. The so-called height function~\cite{Li2022} $h(G, p, k)=Q_{p_1p_2...p_k}(G)$ represents the entanglement entropy between all bipartitions of a graph $G$ respecting a given order $p$ and thus determines the minimum number of required static qubits when creating the graph nodes (photons) in an order $p$~\cite{Li2022}. Since the photons can be emitted in $n!$ different orders, the corresponding height function with the lowest maximum determines the minimum number of static qubits $n_s^{min}(G)$ required to build a target graph state $G$:
\begin{equation}
    n_s^{min}(G)=\min_p \max_k h(G, p, k)
\end{equation}
This quantity corresponds to the linear rank width of the graph~\cite{Li2022, Oum2017}.

When generating a graph state $G$ via one quantum emitter and fusions, $n_s^{min}(G)$ determines a lower bound for the minimum number of fusions $n_f^{min}(G)$. The reason is that a single fusion cannot increase the entanglement entropy between two bipartitions of a graph state by more than one, and we therefore have
\begin{equation}
\label{eq_ne_nf}
    n_f^{min}(G)+1\geq n_s^{min}(G),
\end{equation}
where $+1$ comes from using one static qubit, the quantum emitter, to emit the starting state.

We find that $n_f^{min}=n_s^{min}-1$ for many of the graph states considered here, saturating Eq.~\eqref{eq_ne_nf} (see supplemental material). The smallest graph state for which the inequality is not saturated is the seven-qubit graph state shown in Fig.~\ref{fig:many_fusion_graphs}(c). Despite the correlation between $n_f^{min}$, $n_s^{min}$ for many states, $n_f^{min}\gg n_s^{min}$ is possible for larger graph states. An example is given in Fig.~\ref{fig:many_fusion_graphs}(e) where $n_s^{min}=2$ but $n_f^{min}$ increases by two for every ring attached to the linear chain graph and can thus be arbitrarily large when the graph is continued horizontally. This difference is because the same static qubit can be recycled to open and close every new ring, whereas two new fusions per ring are required. At the same time, we observe that any height function for any ordering of the nodes will at least go up by one step and down by one step (two steps) for every new ring that is added horizontally.

Based on this observation, we conjecture that the minimum number of times the height function climbs up and down above a baseline, set by the number $n_s$ of used static qubits, gives an upper bound for the required number of fusions,
\begin{align}
    \label{eq_climb}
    &n_f^{min}(G, n_s) \leq \text{clmb}(G, n_s),
\end{align}
where the minimum number of climbs $\text{clmb}(G, n_s)$ over all possible permutations, $p$, is given by $\min_p |\{k | h(G,p,k)\neq h(G,p,k+1)\land (h(G,p,k) > n_s \lor h(G,p,k+1) > n_s)\}|$.

For up to 10-qubit graph states, we have checked numerically that the above inequality is fulfilled (assuming $n_s=1$ as done everywhere in this paper). Similar to the linear rank width that determines $n_s^{min}$, it is likely computationally hard to compute $\text{clmb}(G, n_s)$ for large graphs due to the superexponentially increasing number of orderings $p$. Nevertheless, one may take the minimum over a subset of all orderings, which gives an upper bound even without searching through all possibilities.

\subsection{Constructing loss-tolerant graph codes}
\label{sec_codes}
As an example of an application of the lookup table, we use it to search for loss-tolerant graph codes where (a) the encoded quantum information can be recovered with a high tolerance to photon loss and (b) only a small number of fusions are required for their generation. A graph code can be defined by a graph $G=(V, E)$ and a subset of its vertices $I\subseteq V$~\cite{Bell2022}. With $\ket{G}$ being the graph state corresponding to $G$, a logical qubit can be defined by the two graph basis states $\ket{\Bar{0}}=\ket{G}$ and $\ket{\Bar{1}}=Z_I\ket{G}$. Two logical Pauli operators can be defined as $\Bar{X}=\prod_{i\in I}Z_i$ and $\Bar{Z}=S_{i^*}=X_{i^*}\prod_{j\in N(i^*)}Z_j$ ($i^*\in I$) with the stabilizers, $\mathcal{S}$, of the code space being generated by $\{S_i\}_{i\notin I}\cup\{S_i\Bar{Z}\}_{i\in I\setminus i^*}$~\cite{Bell2022}. A measurement-based initialization of a logical qubit can be achieved by starting with a graph state corresponding to a \textit{progenitor graph} $G' = (V + \delta, E + \set{(\delta, i)| i\in I})$ where an additional \textit{encoding qubit} $\delta\notin V$ is connected to all nodes $i\in I$. The logical qubit can then be initialized by measuring the encoding qubit~\cite{Bell2022}.

We search in the lookup table for graph states that make good loss-tolerant codes for logical Pauli measurements. In the presence of physical loss, the logical $\Bar{X} (\Bar{Y}, \Bar{Z})$ Pauli operator remains measurable as long as there is an operator of the form $\Bar{X}S$ ($\Bar{Y}S$, $\Bar{Z}S$) with stabilizers $S\in\mathcal{S}$ that has no support on any physically lost qubit. Otherwise, there is a logical loss. Since $\Bar{X}S$ are all logical $\Bar{X}$ operators, there is some freedom of choice in the used physical measurement basis when such a logical operator is measured by sequentially measuring physical qubits. Given information about previously lost qubits, an adaptive basis choice is used to minimize the probability of logical loss (see~\cite{Bell2022} for details). For a physical loss rate lower than a certain loss threshold, the probability of logical loss will be lower than the physical loss rate\footnote{To ensure first-order loss tolerance, we assume that the encoding qubit cannot be lost. This is justified if the encoding qubit is the spin of a quantum emitter (not the case here) or if the majority of the loss happens in the transmission of the graph code (without sending the encoding qubit).}. The loss threshold can differ depending on which logical Pauli operator is measured, and, in the following, the loss threshold refers to the minimum value for all three logical Pauli measurements.

\begin{figure}[!t]
\includegraphics[width=1.0\linewidth]{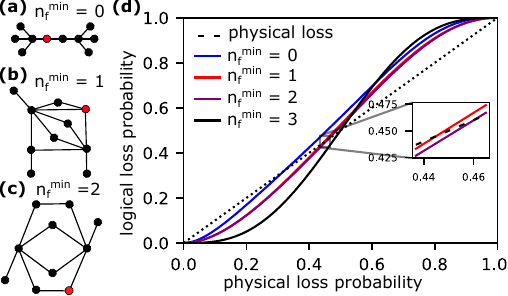}
\caption{\label{fig:codes} \textbf{(a)-(c)} Best graph codes (progenitor graphs) obtained for $n_f^{min}=0$ (a), $n_f^{min}=1$ (b), and $n_f^{min}=3$ (c) fusions, obtained from a search over graphs up to ten nodes. The red nodes indicate the encoding qubit $\delta$, which generates a logical qubit encoded in the graph code upon its measurement~\cite{Bell2022}. \textbf{(d)} Logical loss rate as a function of physical loss rate, where every curve corresponds to the logical Pauli measurement with the lowest loss threshold. Plots are shown for the graph codes from (a, b, c) and the cube graph state representing the best graph code for $n_f^{min}=3$~\cite{Bell2022}.}
\end{figure}

In our search, every graph in the lookup table is the progenitor graph for a graph code (including the encoding qubit), and we loop over all possible encoding qubits. We chose the graph that gives the highest loss threshold for a corresponding choice of the encoding qubit. In Figs.~\ref{fig:codes}(a,b,c), we present the progenitor graph states for the best graph codes that can be obtained by a minimum number ($n_f^{min}$) of zero, one, two, and three fusions. The corresponding loss thresholds are $0.33, 0.45$, $0.46$, and $0.5$ (see Fig.~\ref{fig:codes}(d)). For a loss rate below these thresholds, the transmission probability of an encoded logical qubit exceeds the transmission probability of a physical qubit. All loss thresholds are close to the limit of $0.5$ imposed by the measurement complementarity principle and for $n_f^{min}=3$ fusions, the cube graph state saturates this bound~\cite{Bell2022} (Fig.~\ref{fig:many_fusion_graphs}(d)).

\section{Conclusion and Outlook}
We have presented a dynamic programming approach to build graph states with a single quantum emitter and a minimum number of successful linear optical fusions. Starting from graph states with up to 14 qubits that can be generated with a single quantum emitter, we find constructions of many graph states and graph codes with just a few fusions. Caterpillar tree graph states with up to $14$ qubits~\cite{Thomas2022} have recently been generated by quantum emitters, and fusions have been applied to such states~\cite{Meng2023b, Thomas2024}. Therefore, the found graph state construction protocols may be implemented in the near future. Achieving error and loss rates sufficiently small for photonic quantum applications is still an experimental challenge. However, compared to schemes relying on multiplexed heralded photon sources~\cite{Bartolucci2021b, Maring2024}, the overhead in the number of fusions is strongly reduced. This shows that quantum emitters are a powerful resource for constructing complex photonic states when combined with fusions. Independent of experimental progress, the studied dynamical programming approach can easily be applied to a different class of initial states. Such states could be three-photon star graph states~\cite{Zaidi2015, Lee2023, Gimeno2015, Wiesner2024} which are small enough to be generated in a heralded way by linear optics setups~\cite{Bartolucci2021b, Maring2024}.

We find in total $\sim 2.8\times 10^7$ graph state constructions that are contained in a publicly available lookup table~\cite{erda2024}. The storage space required for the lookup table increases significantly with the size of the considered graph states. Nevertheless, we believe that the size of the initial graph states could be readily enlarged by a few qubits via technical improvements in our implementation. For creating much larger graph states, we anticipate that heuristic methods for their efficient (yet suboptimal) generation will be required. Such methods could be developed using pathfinding algorithms applied to the orbit graph~\cite{Jena2024} or reinforcement learning~\cite{PetterssonToBePublished}. Nevertheless, even such heuristic methods may benefit from a lookup table. One could, for instance, use a divide-and-conquer strategy~\cite{Barrett2005, Duan2005}, where a large graph state is decomposed into several subgraphs for which an optimal construction is found in the lookup table.

Finally, finding more mathematical properties of fusion-based graph-state constructions would be desirable. In this paper, we have given some initial bounds for the required number of successful fusions that are linked to graph-theoretical properties of the target graph state. By more detailed mathematical analysis, it may be possible to derive tighter bounds that can be computed efficiently. Such bounds can guide heuristic search methods for fusion-based constructions of larger graph states.

\section{Acknowledgements}
We are grateful for financial support from Danmarks Grundforskningsfond (DNRF 139, Hy-Q Center for Hybrid Quantum Networks) and Novo Nordisk Foundation (Challenge project “Solid-Q”). S.P. acknowledges funding from the Villum Fonden research grants No.VIL50326 and No.VIL60743, and support from the NNF Quantum Computing Programme. L.D. acknowledges the EPSRC quantum career development grant EP/W028301/1.

\section{Data Availability}
\label{sec_data}
The lookup table presented in this work can be downloaded from Ref.~\cite{erda2024}. The corresponding source code is provided in Ref.~\cite{git2024}.

\bibliography{main.bbl}

\appendix
\onecolumngrid

\section{Graph states that can be generated by a single quantum emitter}
\label{sec_one_emitter}
The scheme by Lindner and Rudolph~\cite{Lindner2009} utilizes a single quantum emitter with a spin together with optical control pulses to generate graph states. The main idea behind this scheme is shown in Fig.~\ref{fig:lookup}(a-c) together with a suitable level structure. In this section, we first review why this scheme allows generating all graph states with a caterpillar graph structure~\cite{Paesani2023}. We then show that no additional states can be generated by applying additional local gates. This justifies starting the computation of the lookup table from all sets of caterpillar graphs.

\subsection{All caterpillar graph states can be generated by the Lindner-Rudolph scheme}
\label{sec_rl}
\begin{lemma}
\label{lemma_caterpillar}
    A single-quantum emitter with a spin can be used to generate any caterpillar graph state of two types~\cite{Lindner2009}: (A) The spin is a neighbor of a qubit at the end of the caterpillar. (B) The spin is at the end of the caterpillar\footnote{The ends of a caterpillar are defined as the endpoints of the longest possible linear chain to which all other nodes have distance one. If there are several possible sets of endpoints with one including the spin $s$, the one with $s$ is chosen.}.
\end{lemma}
This follows by induction: \textit{initial case -- } The single-qubit state $H\ket{0}_s=\ket{+}_s$ is the only graph state with one node and is a caterpillar graph state. \textit{Induction step --} Assume that every $k$-qubit caterpillar graph $\ket{G_k}$ of type A or B can be generated. Such a graph state can be written as $\ket{G_k}=\frac{1}{\sqrt{2}}\left(\ket{0}_s\ket{\Omega}+\ket{1}_sZ_{N(s)}\ket{\Omega}\right)$, where $\ket{\Omega}$ is the induced subgraph when removing qubit $s$, and $N(s)$ is the graph neighborhood of $s$. A suitable photon generation pulse acts as a CNOT gate generating a new photonic qubit that is in state $\ket{0}_n$ if the spin of the emitter is in state $\ket{0}_s$, and a state $\ket{1}_n$ given that the spin of the emitter is in state $\ket{1}_s$ (see Fig.~\ref{fig:lookup}(a), Ref.~\cite{Lindner2009}). This yields the state $\frac{1}{\sqrt{2}}\left(\ket{0}_s\ket{0}_n\ket{\Omega}+\ket{1}_s\ket{1}_nZ_{N(s)}\ket{\Omega}\right)$ and applying a Hadamard gate $H_n$ on the new photonic qubit yields $\frac{1}{\sqrt{2}}\left(\ket{0}_s\ket{+}_n\ket{\Omega}+\ket{1}_s Z_{N(s)+n}(\ket{+}_n\ket{\Omega})\right)$. This state is the original $k$-qubit caterpillar graph $\ket{G_k}$ with the new qubit $n$ attached as a pendant vertex (leaf node) to the spin $s$. The state corresponds to a caterpillar tree with $k+1$ nodes (see Fig.~\ref{fig:lookup}(b)), and given that any caterpillar graph state $\ket{G_k}$ of type A or B can be generated, also every $(k+1)$ node caterpillar state of type A can be generated. If we instead apply the Hadamard gate $H_s$ on the spin, we get $\frac{1}{\sqrt{2}}\left(\ket{0}_n\ket{+}_s\ket{\Omega}+\ket{1}_n Z_{N(s)+s}(\ket{+}_s\ket{\Omega})\right)$. This is the $(k+1)$-qubit caterpillar graph state from before with the labels $s$ and $n$ interchanged, and so $s$ becomes the pendant vertex (see Fig.~\ref{fig:lookup}(c)). Therefore, any $(k+1)$-qubit caterpillar tree graph state of type B can be generated as well. \qedsymbol{}

Intuitively speaking, the gate $H_s$ (applied after emitting a new photon, $n$) extends the length of the central path of the caterpillar graph by $+1$, whereas the gate $H_n$ appends the photon as another pendant vertex of the caterpillar graph state. Any caterpillar graph can be generated by choosing the sequence of operations $H_s$ and $H_n$ applied after the photon generation pulse. An alternative method that captures this intuition is a graphical proof of Lemma~\ref{lemma_caterpillar} using ZX-calculus~\cite{Wetering2020}.

\subsection{All caterpillars and only caterpillars}
\begin{theorem}
\label{theorem_only_caterpillar}
    A single quantum emitter, local unitaries (including non-Clifford), and local measurements can generate (1) all sets of caterpillar graph states and (2) only graph states that are local Clifford equivalent to sets of caterpillar graphs.
\end{theorem}
\textit{Proof $-$} (1) follows from Lemma~\ref{lemma_caterpillar} and the fact that all sets of caterpillars can be obtained by splitting one caterpillar into pieces by $Z$-basis measurements which remove graph nodes~\cite{Hein2006}. (2) can be shown as follows: every graph state generated by a single quantum emitter, single-qubit unitaries, and single-qubit measurements has a height function~\cite{Li2022} that does not exceed $1$ (when, for computing the height function, the graph nodes/photons are ordered corresponding to the physical photon emission). This follows from the entanglement entropy argument in section~\ref{sec_bounds} of the main text. (In this context, we remark that single-qubit measurements create a vertex minor~\cite{Dahlberg2018} and cannot increase the linear rank width~\cite{Oum2017}). The maximum value of this height function gives an upper bound for the linear rank width of the graph, which therefore cannot exceed $1$~\cite{Li2022}. Graphs with linear rank width of at most $1$ are locally equivalent (by a sequence of local graph complementations) to sets of caterpillars~\cite{Bui2013}, and therefore local Clifford equivalent to sets of caterpillars~\cite{Nest2004}. \qedsymbol{}

\section{Local complementation definition and explanation}
\label{sec:LC}
In this section, we define the operation of local complementation and review the Clifford gates corresponding to this operation. Let $G(V, E)$ be a graph with $N(a)$ being the neighborhood of a node $a\in V$ and $G[N(a)]$ being the induced subgraph of $G$ with support on the nodes in $N(a)$. Local complementation on node $a$, $LC(a)$, changes $G$ by inverting the subgraph $G[N(a)]$ to its complement~\cite{Hein2004}
\begin{equation}
    \label{eq_def_lc}
    E \rightarrow E\  \Delta\ \{(i,j)|\ i,j\in N(a)\}, 
\end{equation}
where $\Delta$ is the symmetric difference. Less formally, $LC(a)$ removes every existing edge of $G$ that connects two nodes in $N(a)$ and adds every non-existing edge between two nodes in $N(a)$. It is well-known that this operation can be realized on the corresponding graph state by applying the following combination of single-qubit gates~\cite{Nest2004, Hein2004}:
\begin{equation}
    \label{eq_lc}
    U^{LC}(a) = \sqrt{-iX_a}\prod_{j \in N(a)}\sqrt{iZ_j} = e^{-i\frac{\pi}{4}X_a}
    \prod_{j \in N(a)}e^{i\frac{\pi}{4}Z_j} = R_aZ_aH_aZ_aR_a\prod_{j \in N(a)}\left(\frac{1+i}{\sqrt{2}}R_jZ_j\right),
\end{equation}
with $R=\ket{0}\bra{0}+i\ket{1}\bra{1}$. For completeness, we explain this relation in the following.

Consider the stabilizer generators of the graph state $S_i=X_i\prod_{j \in N(i)}Z_j$, with $N(i)=\{j\in V\setminus i: (i, j) \in E\}$~\cite{Hein2006}. When applying $U_{LC}(a)$, the stabilizers are transformed as $S_i\rightarrow U_{LC}(a)\ S_i\ U_{LC}^{\dagger}(a)$. Since $U_{LC}(a)$ commutes with the generator $S_a$ as well as generators $S_{d}\text{ with } d \notin N(a)$, these stabilizer generators remain unchanged. Stabilizer generators $S_b$, with $b \in N(a)$ are transformed according to:
\begin{align}
    S_b^{'} = e^{-i\frac{\pi}{4}X_a}
    \prod_{j \in N(a)}e^{i\frac{\pi}{4}Z_{j}}\left(X_bZ_a\prod_{j \in N(b) \setminus a} Z_j\right)e^{i\frac{\pi}{4}X_a}\prod_{j \in N(a)}e^{-i\frac{\pi}{4}Z_{j}} \\
    = e^{-i\frac{\pi}{4}X_a}
    e^{i\frac{\pi}{4}Z_b}\left(X_bZ_a\prod_{j \in N(b) \setminus a} Z_j\right)e^{i\frac{\pi}{4}X_a}
    e^{-i\frac{\pi}{4}Z_b}.
\end{align}
Applying the relations
\begin{align}
\label{eq_propagate}
e^{-i\theta Z}X=Xe^{+i\theta Z}, e^{-i\theta X}Z=Ze^{+i\theta X}
\end{align}
and using the convention $Z_Q\equiv \prod_{q\in Q}Z_q$ we obtain
\begin{align}
\label{eq:LC_step_3}
    S_b^{'} = e^{-i\frac{\pi}{2}X_a}
    e^{i\frac{\pi}{2}Z_b}X_bZ_{N(b) \setminus a}Z_a = 
    X_a Z_b X_bZ_{N(b)} = X_a Z_{N(a)} Z_{N(a)} Z_b X_bZ_{N(b)}.
\end{align}
Since stabilizers form a multiplicative group, we can multiply $S_b^{'}$ by  $S_a = X_a Z_{N(a)}$ and we obtain
\begin{align}
    S_b^{'}\rightarrow S_aS_b^{'} = Z_{N(a)} Z_b X_bZ_{N(b)} = 
    X_bZ_{N(b) \Delta N(a) \setminus b}.
\end{align}
This stabilizer transformation corresponds to transforming the subgraph $G[N(a)]$ to its complement.

\section{Justification of the used lookup table approach}
\label{sec_justify}
Our approach to compute a lookup table for optimum graph constructions starts from a caterpillar graph state or several detached caterpillar graph states. Detached caterpillar graph states can be obtained from a single connected caterpillar by applying $Z$-basis measurements. To these initial states, we apply local gates and fusions that measure the parity $X_AZ_B\land Z_AX_B$ upon success. Under the following conditions, this method gives an optimal construction of graph states using single quantum emitters, single-qubit Clifford gates, single-qubit measurements, and fusions: (1) No additional graph orbits are obtained when single-qubit measurements are not performed initially but alternated with fusions and local gates. (2) Adding measurements (single-qubit or fusion) other than $Z$-basis measurements or using more than one type of fusion does not yield any additional graph orbits. In the following subsection~\ref{subsec_order}, we explain why the first condition is fulfilled, and in subsections~\ref{subsec_1_measure},~\ref{subsec_1_fusion}, we explain why the second condition is fulfilled. Subsection~\ref{sec_optimal} summarizes why this implies that the obtained graph state constructions are optimal in the number of fusions. Furthermore, we give another upper bound on the number of fusions for graphs that are not contained in the lookup table.
\label{sec:one}
\begin{figure*}[!t]
\includegraphics[width=1.0\textwidth]{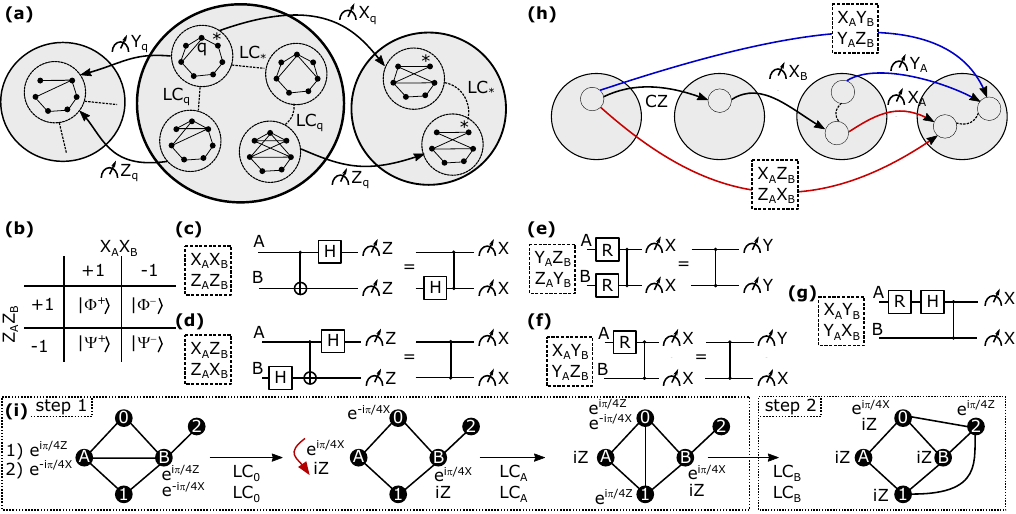}
\caption{\label{fig:one_measurement}\textbf{(a)} Connection between graph orbits induced by different single-qubit measurements from the Pauli group (up to local Clifford gates~\cite{Hein2006}). The illustration shows that it is sufficient to consider $Z$-basis measurements since any state obtained by an $X$- or a $Y$-basis measurement can be reached by applying a $Z$-basis measurement to a different graph state in the orbit. \textbf{(b)} The four Bell states $\ket{\Psi^{\pm}}=1/\sqrt{2}(\ket{01}\pm\ket{10})$, $\ket{\Phi^{\pm}}=1/\sqrt{2}(\ket{00}\pm\ket{11})$ and their stabilizer generators. \textbf{(c)} Circuit implementing the standard Bell state measurement (measuring the parities $X_AX_B\land Z_AZ_B$). \textbf{(d, e, f, g)} Bell state measurement circuit measuring the parities $X_AZ_B\land Z_AX_B$, $Y_AZ_B\land Z_AY_B$, $X_AY_B\land Y_AZ_B$, $X_AY_B\land Y_AX_B$, where we use the Clifford gates $R=\ket{0}\bra{0}+i\ket{1}\bra{1}, H=\frac{1}{\sqrt{2}}(X+Z)$ \textbf{(h)} The parity measurements $X_AZ_B\land Z_AX_B$ and $X_AY_B\land Y_AZ_B$ connect the same graph orbits since their circuits in (d) and (f) only differ by the measurement basis of the last single-qubit measurement, which does not change the orbit according to Lemma~\ref{lemma_single}, as illustrated in subfigure (a). The same argument applies to the parity measurement $Y_AZ_B\land Z_AY_B$. \textbf{(i)} Rewriting a graph state with local Clifford gates applied to it such that only Pauli gates and Clifford gates of type $e^{\pm i\pi/4 Z}$ are applied to the two qubits $A, B$. This is achieved by inserting identities in the form of two identical local complementations. One of the local complementations is applied as a graph transformation, and the other one is applied in the form of the local Clifford gates from  Eq.~\eqref{eq_lc}. Some Pauli operators are propagated (see red arrow for an illustration) to become the last gates applied by flipping the sign of some phases according to Eq.~\eqref{eq_propagate}.}
\end{figure*}
\subsection{Order of the operations}
\label{subsec_order}
Only single-qubit measurements, two-qubit measurements (fusions), and single-qubit gates are used. Thus, it can be assumed that all single-qubit measurements can be performed first. Indeed, the single-qubit measurements commute with the parity measurements corresponding to successful fusions as the latter act on different qubits. Furthermore, any single-qubit Clifford gates (e.g. as part of a local complementation) would only rotate a single-qubit Pauli measurement to a different Pauli measurement and thus would not change the above conclusion leading to Lemma~\ref{lemma_order}.
\begin{lemma}
Assume we can construct a state by local gates, single-qubit measurements, and fusions (Bell state measurements). As the order of measurements is irrelevant, all single-qubit measurements can always be performed before the fusions. \label{lemma_order}
\end{lemma}

\subsection{One single-qubit measurement is sufficient}
\label{subsec_1_measure}
There are three different single-qubit Pauli measurements that project into one of the eigenstates of the Pauli operators $X=\ket{0}\bra{1}+\ket{1}\bra{0}, Y=i\ket{1}\bra{0}-i\ket{0}\bra{1}, Z=\ket{0}\bra{0}-\ket{1}\bra{1}$. Similarly, there are different two-qubit Bell state measurements that measure pairs of parities from the Pauli group. These Bell state measurements can be realized by rotating the standard fusion circuit with Clifford gates~\cite{Gimeno2016, Lobl2024}. Applying different measurements (single-qubit or fusion) to a given graph can lead to different graphs that are not locally equivalent and, thus, are in different graph orbits~\cite{Lobl2024}. However, only the fact that all graph orbits are present in the orbit graph is relevant here. In this case, it is sufficient to consider only one type of single-qubit measurement (say the Pauli $Z$-basis measurement) and one type of fusion. For single-qubit measurements, the reason is Lemma~\ref{lemma_1qbt_measure}.
\begin{lemma}
\label{lemma_single}
A single-qubit Pauli measurement on a graph state creates a state that, up to local Clifford gates, corresponds to a graph state that can be obtained by a $Z$-basis measurement (corresponding to vertex deletion) and local complementations. Up to local Clifford gates, the obtained state thus is a vertex minor of the original graph state~\cite{Dahlberg2018} \label{lemma_1qbt_measure}
\end{lemma}
The proof of Lemma~\ref{lemma_1qbt_measure} is straightforward since, up to some local Clifford gates, (1) a $Z$-measurement corresponds to deleting the measured qubit, (2) a $Y$-measurement can be expressed by local complementation, which does not change the graph orbit, followed by deleting the measured vertex, and (3) an $X$-measurement can be expressed as two local complementations, deleting the measured vertex, followed by another local complementation~\cite{Hein2004, Hein2006}. This is illustrated in Fig.~\ref{fig:one_measurement}(a).\qedsymbol{}

\subsection{One fusion type is sufficient}
\label{subsec_1_fusion}
Finally, we consider the question of why it is sufficient to consider a single kind of type-II fusion. Type-II fusions~\cite{Browne2005, Gimeno2016} destructively measure two qubits and, upon fusion success, correspond to measuring a pair of commuting two-qubit operators. We consider fusions that are rotated by applying local Clifford gates before the measurement setup, in which case the following parities/operators from the Pauli group can be measured~\cite{Lobl2024}:
\begin{table}[h]
\begin{center}
 \begin{tabular}{ |c||c|c|c|c|c| } 
 \hline
 $\sigma^{(1)}_A\sigma^{(2)}_B$ & $X_AX_B$ & $X_AY_B$ & $X_AZ_B$ & $Y_AZ_B$ & $X_AY_B$\\ 
 $\sigma^{(3)}_A\sigma^{(4)}_B$ & $Z_AZ_B$ & $Y_AX_B$ & $Z_AX_B$ & $Z_AY_B$ & $Y_AZ_B$\\ 
 \hline
\end{tabular}
\caption{\label{tab:fusions}Different pairs of parities from the Pauli group that can be measured in a Bell-state measurement. The subscripts $A,B$ label the two fusion-qubits participating in the Bell-state measurement. The superscripts indicate that $\sigma^{(i)}$ are four independent Pauli matrices.}
\end{center}
\end{table}

However, for finding all graph orbits, it is sufficient to apply only a single of these fusion types to graph states in existing graph orbits\footnote{If we would generate orbits including all local Clifford equivalent stabilizer states, it is obvious that one fusion type is sufficient because one could let a Clifford gate that rotates the measurement instead act on the state. However, we only apply fusions to the graph states in the orbit, which is more efficient as much fewer states need to be stored. When all fusion types are allowed, this is sufficient since all stabilizer states are local Clifford equivalent to graph states. What we need to show is that one fusion type is sufficient when using only graph states.}. Here, we chose the fusion that measures $X_AZ_B$ and $Z_AX_B$ upon success. The reason why this restriction works is Theorem~\ref{theorem_one_fusion}.
\begin{theorem}
\label{theorem_one_fusion}
Assume that $\sigma^1_A\sigma^2_B$ and $\sigma^3_A\sigma^4_B$ are two operators with $\sigma^i$ being an element of the Pauli group and that none of the two operators nor their product is a single-qubit gate. Further, assume an arbitrary graph state $\ket{G_1}$ in a graph orbit $O_1$ which becomes a graph state $\ket{G_2}$ (up to local Clifford gates) in a graph orbit $O_2$ when measuring the operators $\sigma^1_A\sigma^2_B$ and $\sigma^3_A\sigma^4_B$. Then, there exist two graph states $\ket{G_1'}\in O_1$ and $\ket{G_2'}\in O_2$ such that $\ket{G_1'}$ becomes $\ket{G_2'}$ (up to local Clifford gates) upon measuring the operators $X_AZ_B$ and $Z_AX_B$.
\end{theorem}

\textit{Proof $-$} The standard Bell state measurement measures $X_AX_B$ and $Z_AZ_B$ as those operators are, up to signs, the stabilizers of the four Bell states ($\ket{\Psi^{\pm}}=1/\sqrt{2}(\ket{01}\pm\ket{10})$, $\ket{\Phi^{\pm}}=1/\sqrt{2}(\ket{00}\pm\ket{11})$) as illustrated in Fig.~\ref{fig:one_measurement}(b). In a circuit-based picture, this measurement is implemented by a CNOT gate, followed by an $H$-gate and two $Z$-basis measurements as shown in the left side of Fig.~\ref{fig:one_measurement}(c). To show the equivalence of different Bell-state measurements, it is convenient to transform the CNOT gate into a controlled-$Z$ gate (which transforms one graph state into another one) and single-qubit gates/measurements as illustrated in Fig.~\ref{fig:one_measurement}(c) on the right. The circuit measuring the parities $X_AZ_B$ and $Z_AX_B$ can be obtained by applying an $H$-gate before this circuit, yielding a circuit consisting of just a controlled $Z$-gate and two $X$-basis measurements (see Fig.~\ref{fig:one_measurement}(d)). Applying another Clifford gate $R=\ket{0}\bra{0}+i\ket{1}\bra{1}$ on qubit $A$ before the fusion rotates the measured parities to $X_AY_B\land Y_AZ_B$, and applying $S_AS_B$ rotates them to $Y_AZ_B\land Z_AY_B$. In both cases, $R$-gates commute with the controlled-$Z$ gate and can thus be moved to the right of it, only rotating the final single-qubit measurements as illustrated in Figs.~\ref{fig:one_measurement}(e, f).

We now use the circuit representations to show the equivalence of all five double-parity measurements for constructing the lookup table. Since the final circuits in Fig.~\ref{fig:one_measurement}(d, e, f) only differ by their single qubit measurements, applying Lemma~\ref{lemma_1qbt_measure} twice implies that the parity measurements $Y_AZ_B\land Z_AY_B$, $X_AY_B\land Y_AZ_B$, and $X_AZ_B\land Z_AX_B$ all connect the same graph orbits. Thus, only one of these three double-parity measurements needs to be considered. Fig.~\ref{fig:one_measurement}(h) illustrates the equivalence of the fusions measuring $X_AZ_B\land Z_AX_B$ and $X_AY_B\land Y_AZ_B$. The situation is less obvious for the two other pairs of parity measurements, $X_AX_B\land Z_AZ_B$ and $X_AY_B\land Y_AX_B$, where an $H$-gate is applied before the controlled-$Z$ gate in the corresponding fusion circuit (see Figs.~\ref{fig:one_measurement}(c, g)). We consider these two cases in the following.

First note that the local graph complementation on qubit $q$ corresponds to the operation $\exp\left(-i\frac{\pi}{4}X_q\right)\exp\left(i\frac{\pi}{4}Z\right)_{N(q)}$~\cite{Hein2004, Hein2006} (see Eq.~\eqref{eq_lc}). Therefore, local complementation on qubit $A$ applies the gate $\exp\left(-i\frac{\pi}{4}X_A\right)$, and local complementation on one of its neighbors applies the gate $\exp\left(i\frac{\pi}{4}Z_A\right)$. Furthermore, these two gates generate the single-qubit Clifford group, and so any Clifford gate before the controlled-$Z$ gate on the right-hand side of Figs.~\ref{fig:one_measurement}(c,g) can be expressed as a sequence of these two gates. The key idea is to remove all gates of the type $\exp\left(\pm i\frac{\pi}{4}X\right)$ from this gate sequence by inserting several identities of two identical local complementations $LC_iLC_i$ as illustrated in Fig.~\ref{fig:one_measurement}(i). The first of these two local complementations is expressed as a graph transformation, and the second one is expressed by the gates $\exp\left(-i\frac{\pi}{4}X\right)$ and $\exp\left(i\frac{\pi}{4}Z\right)$. The local complementation identities can be chosen such that these gates cancel all terms except Pauli gates\footnote{Pauli gates can be propagated to the right of the other gates using Eq.~\eqref{eq_propagate}. The Pauli gates could also be ignored as they correspond to stabilizer signs which can be tracked independenly~\cite{Gottesman1998} and so leave the graph orbits invariant.} and the gates $\exp\left(\pm i\frac{\pi}{4}Z\right)$. In particular, one first removes all gates $\exp\left(\pm i\frac{\pi}{4}X_A\right)$ on fusion qubit $A$ (step 1 in Fig.~\ref{fig:one_measurement}(i)) and then all gates $\exp\left(\pm i\frac{\pi}{4}X_B\right)$ on fusion qubit $B$ (step 2 in Fig.~\ref{fig:one_measurement}(i))\footnote{This works if none of the fusion qubits $A, B$ is detached and if the two qubits do not form an isolated two-node graph. Furthermore, step 1 needs to be applied to qubit $B$ in case it is only connected to $A$. In the other boundary cases, successful fusion anyway has the effect of single-qubit measurements~\cite{Lobl2024} and, therefore, these cases can be disregarded.}.

The described method can be applied to the circuits on the right-hand side of Figs.~\ref{fig:one_measurement}(c, g). In Fig.~\ref{fig:one_measurement}(c), the $H$-gate on the lower qubit can be expressed as $H=-i \exp\left(i\frac{\pi}{4}Z\right)\exp\left(i\frac{\pi}{4}X\right)\exp\left(i\frac{\pi}{4}Z\right)$. Where the $\exp\left(i\frac{\pi}{4}Z\right)$ gate that is applied last commutes with the $CZ$-gate and can therefore be converted into a change of the single qubit measurement operator. The remaining gates are $\exp\left(i\frac{\pi}{4}X\right)\exp\left(i\frac{\pi}{4}Z\right)$ where $\exp\left(i\frac{\pi}{4}X\right)$ does not commute with the $CZ$-gate. First, we need to remove the $\exp\left(i\frac{\pi}{4}Z\right)$ gate by inserting two local complementations on a neighboring qubit. One local complementation does the graph transformation and the other one cancels the gate. Then, the term $\exp\left(i\frac{\pi}{4}X\right)$ can be removed by inserting an identity of local complementation on the qubit itself. In Fig.~\ref{fig:one_measurement}(g), the gate on the upper qubit can be expressed as $RH=\frac{1-i}{\sqrt{2}}\exp\left(i\frac{\pi}{4}X\right)\exp\left(i\frac{\pi}{4}Z\right)$ and thus can be removed with the same procedure. After these steps, all gates on fusion qubits that remain on the left-hand side of the $CZ$-gate commute with it (except the Pauli $X, Y$ gates, which can be propagated to the right of the $CZ$-gate by using the relation $X_i CZ_{i,j} = CZ_{i,j}X_iZ_j$).

Consequently, the circuits in Figs.~\ref{fig:one_measurement}(c, g) can be converted into circuits where the $CZ$-gate is applied first, followed by rotated single-qubits measurements of the fusion qubits $A, B$ and, potentially, some Clifford gates on other qubits. Since the gates on other qubits commute with the fusion and do not change the graph orbit, applying Lemma~\ref{lemma_single} twice implies that also the parity measurements $X_AX_B\land Z_AZ_B$ or $X_AY_B\land Y_AX_B$ do not yield any additional graph orbits. \qedsymbol{}

\subsection{Optimality of the constructions}
\label{sec_optimal}
An important result regarding the optimality of our constructions follows from an argument already used in the proof of Theorem~\ref{theorem_only_caterpillar}. The graph-state construction starts from sets of caterpillar tree graphs which have linear rank width $1$~\cite{Bui2013}. All single-qubit Pauli measurements can be performed first (Lemma~\ref{lemma_order}) and generate a state that, up to local Clifford gates, is a vertex minor of the graph (Lemma~\ref{lemma_1qbt_measure}). Since the linear rank width can only decrease upon taking a vertex minor~\cite{Oum2017}, the resulting state is, up to local Clifford gates, a graph with linear rank width $1$ or less. Such a graph is locally equivalent (by a sequence of local graph complementations) to a set of caterpillars~\cite{Bui2013}. Therefore, the state resulting from the single-qubit measurement is local Clifford equivalent to caterpillar graph states~\cite{Nest2004}. Since we already start from all caterpillar graph states up to a certain size, applying single-qubit Pauli measurements is thus no additional resource.

Now, assume we would allow for larger initial states than the maximum of 14 qubits in our lookup table. Any construction of a target state that we already found cannot be improved using a larger initial state since we would either need to add fusions or single-qubit measurements to reach the number of qubits in the target state. Adding single-qubit measurements would be redundant as we could have started with a corresponding set of caterpillars with fewer nodes, which would be part of the computed lookup table. This leads to Theorem~\ref{theorem_optimality}.
\begin{theorem}
    \label{theorem_optimality}
    Let $f_n(X)$ be a function that gives the minimum number of fusions required to construct a graph state $X$ by applying fusions to a graph state with $n$ or less qubits that a single quantum emitter can generate ($f_n(X)\equiv-1$ if no such construction exists), then: $f_n(X)\geq 0\Rightarrow f_n(X)=f_{\infty}(X)$.
\end{theorem}
In other words, any fusion-based graph state construction, which is optimum in the number of fusions when allowing arbitrary sets of caterpillar trees up to a fixed number $n$ of nodes/qubits as initial states, is also optimal in the number of fusions when $n$ is larger.

Along the same lines, one can use the lookup table to estimate a lower bound for the number of fusions even when a state $\ket{G}$ is not in the lookup table. Assume one can get to $\ket{G}$ by taking the state $\ket{K}$, for which the lookup table gives an optimum construction with $n$ fusions, and adding a new node by connecting it to one or more vertices in $\ket{K}$. If there was a construction of $\ket{G}$ with less than $n$ fusions, the construction of $\ket{K}$ in the lookup table could not be optimal: one could instead take the construction of $\ket{G}$ with fewer fusions, delete the additional node by a $Z$-basis measurement, yielding a better construction of $\ket{K}$. This leads to Theorem~\ref{theorem_subgraph} saying that the subgraph that is most difficult to generate determines a lower bound for the required number of fusions.
\begin{theorem}
    \label{theorem_subgraph}
    Let $f(G)$ be a function that gives the minimum number of fusions required to construct a graph state $G$ via one quantum emitter and fusions, and let $\mathcal{H}(G)$ be a set of graphs that are induced subgraphs of a graph $G$, then: $f(G) \geq \text{max}_{X\in \mathcal{H}(G)} f(X)$. 
\end{theorem}
Using Theorem~\ref{theorem_subgraph}, we can, for instance, say that constructing any of the graph states in the lower part of Fig. 5(a) from Ref.~\cite{Bell2022} requires at least three fusions since all of these graphs have the cube as a subgraph for which three fusions are required.

\section{Further graph constructions}
\label{sec_constr}
In this section, we show fusion-based constructions for all graph states from the main text. These constructions are shown in Table~\ref{table_construct}. First, we provide the constructions for the graph states from Fig.~\ref{fig:many_fusion_graphs}(b-d). One of these graph states is the cube graph for which an alternative construction using different types of fusions has been presented in Ref.~\cite{Lobl2024}. Furthermore, we provide the constructions of the two graph codes from Figs.~\ref{fig:codes}(b,c) as well as the $N=4$ repeater graph state. For the latter, we need one fusion, which is one fusion less than a suggestion in Ref.~\cite{Buterakos2017}. Interestingly, these repeater graph states are the only examples for which our optimum construction starts from detached caterpillar graphs (also in Fig.~\ref{fig:examples}(b)). Finally, we consider a class of graph states from Ref.~\cite{Morley2019} where loss-tolerant teleportation has been studied. In particular, we consider so-called crazy graphs with an additional input and output node. These graphs have $k$ layers of $m$ nodes where two neighboring layers form a complete bipartite graph. An example of a construction of such graph states (with $k=m=3$) is shown in Fig~\ref{fig:examples}(a) of the main text. We find that all graphs with $k=1$ and $k=2$ can be generated from a single quantum emitter and local complementations. Fusions are thus only required for $k>2$. The crazy graph construction for $k=2$ is shown at the bottom of Table~\ref{tab:fusions}.

\begin{table}[!h]
\caption{Fusion-based construction of several interesting graph states. The operations to turn the initial graph states into the target graph state are applied from left (first) to right (last).}
\label{tabe_construction}
\centering
\label{table_construct}
\begin{tabular}{ |c|c|c|c| }
 \hline
 \thead{source} & \thead{initial graph} & \thead{operations} & \thead{target graph} \\
 \hline
 \hline
 Fig.~\text{\ref{fig:many_fusion_graphs}}(b), No. 19~\cite{Hein2004} & \makecell{\includegraphics[width=0.24\textwidth]{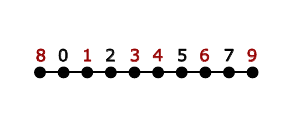}} & \makecell{$LC(0)$, fuse $0\& 5$, $LC(1)$,\\ fuse $2\& 7$, $LC(1)$} & \makecell{\includegraphics[width=0.12\textwidth]{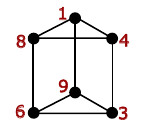}} \\
 \hline
 Fig.~\text{\ref{fig:many_fusion_graphs}}(d), No. 141~\cite{Cabello2009} & \makecell{\includegraphics[width=0.24\textwidth]{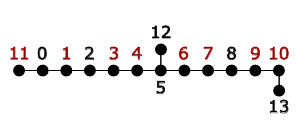}} & \makecell{fuse $0\& 5$, fuse $12\& 13$,\\ $LC(1), LC(2), LC(1)$, fuse $2\& 8$} & \makecell{\includegraphics[width=0.12\textwidth]{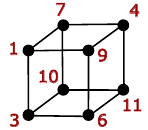}}\\
 \hline
 Fig.~\text{\ref{fig:many_fusion_graphs}}(d), No. 142~\cite{Cabello2009} & \makecell{\includegraphics[width=0.24\textwidth]{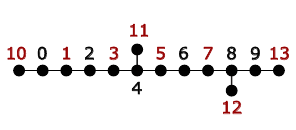}} & \makecell{fuse $4\& 8$, $LC(0)$, fuse $0\& 6$, $LC(1)$,\\ fuse $2\& 9$, $LC(7)$, $LC(11)$, $LC(12)$} & \makecell{\includegraphics[width=0.12\textwidth]{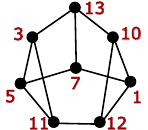}}\\
  \hline
 Fig.~\text{\ref{fig:many_fusion_graphs}}(d), No. 144~\cite{Cabello2009} & \makecell{\includegraphics[width=0.24\textwidth]{gr2_4_5_init.pdf}} & \makecell{fuse $0\& 5$, fuse $12\& 13$,\\ $LC(1), LC(2)$, fuse $2\& 8$, $LC(1)$} & \makecell{\includegraphics[width=0.12\textwidth]{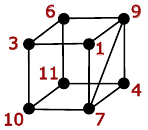}} \\
 \hline
 Fig.~\text{\ref{fig:many_fusion_graphs}}(d), No. 145~\cite{Cabello2009} & \makecell{\includegraphics[width=0.24\textwidth]{gr2_4_5_init.pdf}} & \makecell{fuse $0\& 5$, fuse $2\& 8$, fuse $12\& 13$} & \makecell{\includegraphics[width=0.12\textwidth]{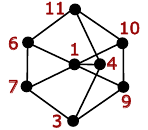}} \\
 \hline
 Fig.~\text{\ref{fig:many_fusion_graphs}}(d), No. 146~\cite{Cabello2009} & \makecell{\includegraphics[width=0.24\textwidth]{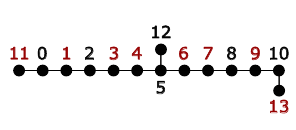}} & \makecell{fuse $0\& 5$, fuse $2\& 8$, $LC(1)$,\\ fuse $10\& 12$, $LC(1)$} & \makecell{\includegraphics[width=0.12\textwidth]{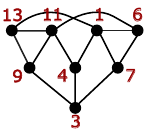}} \\
 \hline
 Fig.~\text{\ref{fig:codes}}(b) & \makecell{\includegraphics[width=0.24\textwidth]{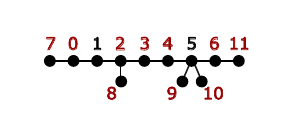}} & fuse $1\& 5$ & \makecell{\includegraphics[width=0.12\textwidth]{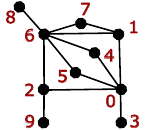}} \\
 \hline
 Fig.~\text{\ref{fig:codes}}(c) & \makecell{\includegraphics[width=0.24\textwidth]{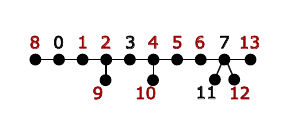}} & \makecell{fuse $0\& 7$, $LC(1), LC(4)$,\\ fuse $3\& 11$, $LC(4), LC(2), LC(12)$} & \makecell{\includegraphics[width=0.12\textwidth]{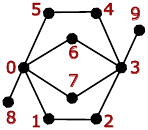}} \\
 \hline
 $N=4$ repeater & \makecell{\includegraphics[width=0.24\textwidth]{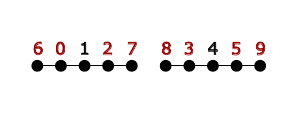}} & \makecell{$LC(1), LC(0)$, fuse $1\& 4$, $LC(0)$} & \makecell{\includegraphics[width=0.12\textwidth]{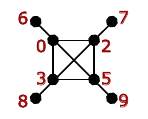}} \\
 \hline
 $k=2$ crazy & \makecell{\includegraphics[width=0.24\textwidth]{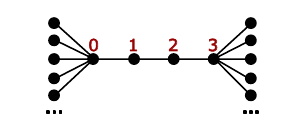}} & \makecell{$LC(0), LC(1), LC(2), LC(3), LC(0)$} & \makecell{\includegraphics[width=0.12\textwidth]{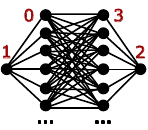}} \\
 \hline
\end{tabular}
\end{table}

\end{document}